\documentclass[adp,fleqn]{w-art}
\usepackage{amsfonts}
\usepackage{amsmath}
\usepackage{times}
\usepackage{w-thm}
\usepackage{graphicx}

\chardef\bslash=`\\

\begin{document}

\title[BF-DMFT]{Mixtures of correlated bosons and
fermions: Dynamical\\ mean-field theory for normal and
condensed phases}
\author[K.\ Byczuk]{Krzysztof Byczuk\thanks{%
Corresponding author \quad E-mail: \textsf{byczuk@fuw.edu.pl}, Phone:
+48\,22\,55 32\,256, Fax: +48\,22\,6219\,475}\inst{1,2} \and Dieter Vollhardt%
\thanks{%
Dieter.Vollhardt@Physik.Uni-Augsburg.de}\inst{2}}

\begin{abstract}
We derive a dynamical mean-field theory for mixtures of interacting
bosons and fermions on a lattice (BF-DMFT). The BF-DMFT is a
comprehensive, thermodynamically consistent framework for the
theoretical investigation of Bose-Fermi mixtures and is applicable
for arbitrary values of the coupling parameters and temperatures. It
becomes exact in the limit of high spatial dimensions $d$ or
coordination number $Z$ of the lattice. In particular, the BF-DMFT
treats normal and condensed bosons on equal footing and thus
includes the effects caused by their dynamic coupling. Using the
BF-DMFT we investigate two different interaction models of
correlated lattice bosons and fermions, one where all particles are
spinless (model I) and one where fermions carry a spin one-half
(model II). In model I the local, repulsive interaction between
bosons and fermions can give rise to an attractive  effective
interaction between the bosons. In model II it can also lead to an
attraction between the fermions.

\end{abstract}

\maketitle

\DOIsuffix{theDOIsuffix} %%
\Volume{12} \Issue{1} \Copyrightissue{01} \Month{01} \Year{2003} %%
\pagespan{1}{} %%
\Receiveddate{15 November 1900} %%
\Accepteddate{2 December 1900} %% \Dateposted{3 December 1900}

\address[\inst{1}]{Institute of Theoretical Physics, University of Warsaw,
ul. Ho\.za 69, 00-681 Warszawa, Poland} %%

\address[\inst{2}]{Theoretical Physics III, Center for Electronic
Correlations and Magnetism, Institute for Physics, University of Augsburg,
86135 Augsburg, Germany} %%

\section{Introduction}

Ultracold atoms in optical lattices are not only fascinating
many-body systems \textit{per se}, they  also help to better
understand interaction models and correlation phenomena in condensed
matter physics \cite{lewenstein06,bloch08}. Indeed, the high
tunability of experimental parameters in these systems allows to
explore the properties of correlated fermions and bosons in
parameter regimes which are not accessible in condensed matter
systems provided by nature \cite{bloch08}. For example, precise
measurements of the effects of strong disorder on interacting bosons
have now become possible \cite{white09}. Similarly, important
insights can be expected from investigations of strongly disordered
interacting fermions, for which detailed predictions exist
\cite{falko09,aguiar09,byczuk05,byczuk09,foster08,paris07,denteneer01}. Ultracold
atoms in optical lattices may also be employed to realize entirely
new physical systems, e.g.,
 mixtures with an arbitrary relative concentration of lattice fermions and bosons
% and where number of bosons is conserved
 \cite{gunter06,ospelkaus06,best09}.

Quantum many-body problems in the thermodynamic limit can almost
never be solved exactly and thus
require approximate investigation methods. During the last 20 years
the dynamical mean-field theory (DMFT) has proved to be a reliable
approximation scheme for correlated
lattice fermions (electrons) in dimension $d=3$ \cite%
{metzner89,pruschke95,georges96,kotliar04}. The bosonic counterpart
of the DMFT --- the B-DMFT --- was developed only most recently
\cite{byczuk08}. The B-DMFT is a thermodynamically consistent,
conserving and non-perturbative theoretical framework for the
investigation of correlated bosons on a lattice. It is applicable
for all physical parameters and becomes exact in the limit of high
spatial dimensions $d$ or lattice coordination number $Z$. The
B-DMFT has the important property that it treats normal and
condensed bosons on equal footing and thus includes the effects
caused by their dynamic coupling \cite{byczuk08}. This is in
contrast to the seminal static mean-field theory of Fisher \emph{et
al.} \cite{fisher89}, and the equivalent Gutzwiller approximation
\cite{rok}, which focus on the condensed bosons and treat normal
bosons as immobile.
Explicit results of the B-DMFT were so far obtained for the bosonic
Falicov-Kimball model \cite{byczuk08} and the bosonic Hubbard model
\cite{byczuk08,hubner09,hu09,semerjian09}. In this paper we extend
the DMFT and B-DMFT and  formulate a dynamical mean-field theory for
\emph{mixtures} of correlated lattice bosons and fermions. The
theoretical framework derived in this way, referred to as BF-DMFT in
the following, is a generalization of both the fermionic DMFT and
the B-DMFT. It provides a set of self-consistency equations which
are valid for arbitrary coupling parameters and temperatures, and
which describe the static and dynamic properties of a system of
mutually interacting lattice bosons and fermions. These BF-DMFT
equations are exact in the limit $Z\rightarrow \infty$ and provide a
comprehensive, non-perturbative approximation scheme for Bose-Fermi
 mixtures on lattices with
finite $Z$. Similar to the B-DMFT the BF-DMFT derived here treats
normal and (pair-)condensed particles on equal footing.

Since the BF-DMFT includes many-body correlations and the dynamic
coupling between normal and condensed particles in a systematic way it
extends earlier theoretical investigations of
Bose-Fermi mixtures formulated in the continuum
\cite{viverit00,bijlsma00,albus02,chui104,rothel07,enss08} and on
lattices
\cite{buchler03,buchler04,lewenstein04,snoek05,dickerscheid05,ahufinger05,wang05,sanchez06,sengupta07,lutchyn08,refael08,pollet08,titvinidze08,tewari09,lutchyn09}.
A dynamic coupling between fermions in Bose-Fermi mixtures was included by Titvinidze, Snoek, and Hofstetter \cite{titvinidze08} who treated fermions within DMFT and bosons within the Gutzwiller approximation.
 One of the most interesting questions in this context concerns
the renormalization of the interaction between bosons and between
fermions due to the interaction between bosons and fermions.
Employing the many-body BF-DMFT formulated in this paper we find
that the effective interactions between bosons and between fermions
can indeed be \emph{attractive}, and can thus lead to phase
instabilities towards superfluid fermions or phase-separated bosons.
This confirms and extends the range of validity of earlier
investigations performed within a T-matrix approximation or the
static mean-field Bogoliubov approximation
\cite{viverit00,bijlsma00,albus02,buchler03,buchler04,chui104,wang05,rothel07,sengupta07,enss08,lutchyn08,refael08,pollet08,titvinidze08,lutchyn09}.

In section 2 we introduce two models of correlated, spinless lattice
bosons interacting with correlated lattice fermions without spin
(model I) and with spin (model II). In sections 3 and 4 we derive
the BF-DMFT self-consistency equations for these models
and discuss some of their properties, e.g., the emergence of
effective attractions between bosons due to the interaction with
fermions, and \emph{vice versa}.

\section{Lattice models for mixtures of correlated bosons and fermions}

In the following we consider two different models for correlated lattice
bosons ($b$) and fermions ($f$) with local interactions. In the first model
(model I) all particles or atoms are spinless. Such a system can be obtained
experimentally by preparing a mixture where $^{40}$K fermionic and $^{87}$Rb
bosonic atoms are in only one hyper-fine state \cite%
{gunter06,ospelkaus06,best09}.
This would correspond to the simplest Bose-Fermi mixture where all
spin-dependent effects are absent.
The Hamiltonian has the form
\begin{eqnarray}
H_{\mathrm{I}}=\sum_{ij}t_{ij}^{b}b_{i}^{\dagger
}b_{j}+\sum_{i}\epsilon _{i}^{b}n_{i}^{b}+\frac{U_{b}}{2}\sum
n_{i}^{b}(n_{i}^{b}-1)+\sum_{ij}t_{ij}^{f}f_{i}^{\dagger
}f_{j}+\sum_{i}\epsilon
_{i}^{f}n_{i}^{f}+U_{bf}\sum_{i}n_{i}^{b}n_{i}^{f}, \label{model_I}
\end{eqnarray}%
where $b_{i}$, $f_{i}$ ($b_{i}^{\dagger }$, $f_{i}^{\dagger }$) are
annihilation (creation) operators on lattice site $i$ for bosons and
fermions, respectively, with number operators
$n_{i}^{b}=b_{i}^{\dagger
}b_{i}$ and $n_{i}^{f}=f_{i}^{\dagger }f_{i}$ and hopping amplitudes $%
t_{ij}^{b}$ and $t_{ij}^{f}$. The interaction between bosons is denoted by $%
U_{b}$, and that between bosons and a fermion by $U_{bf}$. % if they
%are on the same lattice site.
In model I there is no local fermion--fermion interaction because of the
exclusion principle \cite{NN_interactions}. For generality we include local
potentials $\epsilon _{i}^{b}$, $\epsilon _{i}^{f}$, acting on bosons and
fermions, respectively. They can either be applied externally such as by a
harmonic trap, or act intrinsically as in the case of an additional random
potential.

In model II the bosons are still spinless, but the fermions now carry
a spin one-half. Experimentally the $^{40}$K fermionic atoms
should then be prepared in two hyperstates with equal population, whereas the $%
^{87}$Rb bosonic atoms should remain in one hyperfine state. In such
a situation superfluidity of condensed Cooper pairs also becomes
possible. In this case the Hamiltonian takes the form
\begin{eqnarray}
H_{\mathrm{II}} &=&\sum_{ij}t_{ij}^{b}b_{i}^{\dagger }b_{j}+\sum_{i}\epsilon
_{i}^{b}n_{i}^{b}+\frac{U_{b}}{2}\sum n_{i}^{b}(n_{i}^{b}-1)  \nonumber \\
&&+\sum_{ij\sigma }t_{ij}^{f}f_{i\sigma }^{\dagger }f_{j\sigma
}+\sum_{i\sigma }\epsilon _{i}^{f}n_{i\sigma }^{f}+\frac{U_{f}}{2}%
\sum_{\sigma }n_{i\sigma }^{f}n_{i\bar{\sigma}}^{f}+U_{bf}\sum_{i\sigma
}n_{i}^{b}n_{i\sigma }^{f},  \label{model_II}
\end{eqnarray}%
where $\sigma $ denotes the spin (or hyperfine state) of the
fermions. In contrast to (1)  model II contains also a local
interaction $U_{f}$ between fermions with the opposite spins.

We will now derive the self-consistent DMFT equations for model I in
real space \cite{snoek08}. To this end we employ the cavity method \cite{georges96}
together with a scaling of the hopping amplitudes for fermions
\cite{metzner89} and bosons \cite{byczuk08,Bosonic_scaling}.

\section{Solution of model I within Dynamical Mean-Field Theory}

In the DMFT approach the local, site-diagonal Green functions are of
particular interest. The local one-particle bosonic Green functions
are defined as
\begin{equation}
\mathbf{G}_{i}^{b}(\tau )\equiv -\langle T_{\tau }\mathbf{b}_{i}(\tau )%
\mathbf{b}_{i}^{\dagger }(0)\rangle _{S_{i}}\equiv -\left(
\begin{array}{cc}
\langle T_{\tau }b_{i}(\tau )b_{i}^{\dagger }(0)\rangle _{S_{i}} & \langle
T_{\tau }b_{i}(\tau )b_{i}(0)\rangle _{S_{i}} \\
\langle T_{\tau }b_{i}^{\dagger }(\tau )b_{i}^{\dagger }(0)\rangle _{S_{i}}
& \langle T_{\tau }b_{i}^{\dagger }(\tau )b_{i}(0)\rangle _{S_{i}}%
\end{array}%
\right) ,  \label{green_b_loc}
\end{equation}%
where we use the Nambu notation to incorporate the off-diagonal, anomalous
Green functions in the case of Bose-Einstein condensed bosons \cite{rickayzen,agd}, and $T_{\tau }$ represents  the time ordering operator \cite{comment_a}. The
Nambu spinor operators for bosons are defined by
\begin{equation}
\mathbf{b}_{i}=\left(
\begin{array}{c}
b_{i} \\
b_{i}^{\dagger }%
\end{array}%
\right) \;\;\;\mathrm{and}\;\;\;\mathbf{b}_{i}^{\dagger }=(b_{i}^{\dagger
},b_{i}).
\end{equation}%
To characterize the bosons completely we also have to consider the order
parameter of the BEC phase, i.e., the condensate wave function, given by
\begin{equation}
\mathbf{\Phi }_{i}(\tau )\equiv \langle \mathbf{b}_{i}(\tau )\rangle
_{S_{i}}.
\end{equation}%
The square of the absolute value of $\mathbf{\Phi }_{i}(\tau )$ yields the
number of bosons in the condensate at site $i$, i.e.
\begin{equation}
n_{i}^{\mathrm{BEC}}=\langle b_{i}^{\dagger }(\tau )\rangle _{S_{i}}\langle
b_{i}(\tau )\rangle _{S_{i}}=\frac{1}{2}|\mathbf{\Phi }_{i}(\tau )|^{2}.
\end{equation}%
The equilibrium condensate density is expected to be independent of
time. This still allows for a time-dependence of the phase of
$\mathbf{\Phi }_{i}(\tau )$. The local, site diagonal one-particle
fermionic Green function is defined by
\begin{equation}
G_{i}^{f}(\tau )\equiv -\langle T_{\tau }f_{i}(\tau )f_{i}^{\dagger
}(0)\rangle _{S_{i}}.  \label{green_f_loc}
\end{equation}%
The local functions defined above are determined by the local action $S_{i}$
which will be derived next.

Using the path integral formalism \cite{negele} the exact grand
canonical partition function for the lattice Bose-Fermi mixture is
obtained from
\begin{equation}
Z=\int D[b]\int D[f]e^{-S[b,f]},
\end{equation}%
where the action $S[b,f]$ is given by
\begin{equation}
S[b,f]=\int_{0}^{\beta }d\tau \left\{ \sum_{i}\left[ b_{i}^{\ast }(\tau
)(\partial _{\tau }-\mu _{b})b_{i}(\tau )+f_{i}^{\ast }(\tau )(\partial
_{\tau }-\mu _{f})f_{i}(\tau )\right] +H_{\mathrm{I}}[b,f]\right\} .
\end{equation}%
Here $\mu _{b}$ and $\mu _{f}$ are the chemical potentials of bosons
and fermions, respectively, and $\beta =1/k_{B}T$ is the inverse
temperature. In the cavity method \cite{georges96} we select a
single lattice site $i_{0}$ and split $S[b,f]$ into three
contributions:
\begin{equation}
S[b,f]=S_{i_{0}}[b,f]+\Delta S_{i_{0}}[b,f]+S_{i\neq i_{0}}[b,f].
\end{equation}%
Here the first term,
\begin{eqnarray}
S_{i_{0}}[b,f] &=&\int_{0}^{\beta }d\tau \left\{ b_{i_{0}}^{\ast }(\tau
)(\partial _{\tau }-\mu _{b}+\epsilon _{i_{0}}^{b})b_{i_{0}}(\tau
)+f_{i_{0}}^{\ast }(\tau )(\partial _{\tau }-\mu _{f}+\epsilon
_{i_{0}}^{f})f_{i_{0}}(\tau )\right.  \nonumber \\
&&\left. +\frac{U_{b}}{2}n_{i_{0}}^{b}(\tau )(n_{i_{0}}^{b}(\tau
)-1)+U_{bf}n_{i_{0}}^{b}(\tau )n_{i_{0}}^{f}(\tau )\right\} ,
\end{eqnarray}%
corresponds to the action for a separated site $i_{0}$. The second
term,
\begin{equation}
\Delta S_{i_{0}}[b,f]=\int_{0}^{\beta }d\tau \sum_{j\neq
i_{0}}(t_{i_{0}j}^{b}b_{i_{0}}^{\ast }b_{j}+t_{ji_{0}}^{b}b_{j}^{\ast
}b_{i_{0}}+t_{i_{0}j}^{f}f_{i_{0}}^{\ast }f_{j}+t_{ji_{0}}^{f}f_{j}^{\ast
}f_{i_{0}}),
\end{equation}
is due to hopping processes between site $i_{0}$ and all other sites, and
the third term, $S_{i\neq i_{0}}[b,f]$, represents the remaining parts of
the action. In the next step we expand the exponential function with respect
to $\Delta S_{i_{0}}[b,f]$ and perform the functional integrals with respect
to the complex variables $b_{i}$ and Grassmann variables $f_{i}$ with $i\neq
i_{0}$. The resulting expression for the partition function contains
infinitely many correlation functions with different numbers of operators.
Within DMFT we need to keep only terms with a single $b_{i_{0}}$ operator as
well as pairs of bosonic or fermionic operators. These terms are then
re-exponentiated according to the linked cluster theorem \cite%
{georges96,byczuk08}. This approximation becomes exact in the limit
of infinite dimensions $d$ or coordination numbers $Z$, provided the
hopping amplitudes for fermions, $t_{ij}^{f}$, \cite{metzner89} and
bosons, $t_{ij}^{b}$,  \cite{byczuk08,Bosonic_scaling} are
appropriately scaled with $d$ or$Z$. In this limit the self-energies
are diagonal in lattice indices, i.e., $\Sigma _{ij}^{b,f}(\tau
)=\Sigma _{i}^{b,f}(\tau )\delta _{ij}$. As a result one obtains a
sum of three local DMFT actions for the lattice site $i_{0}$ as
\begin{equation}
S_{i_{0}}=S_{i_{0}}^{\mathrm{b}}+S_{i_{0}}^{\mathrm{f}}+S_{i_{0}}^{\mathrm{bf%
}}.
\end{equation}%
Here
\begin{eqnarray}
S_{i_{0}}^{\mathrm{b}} &=&\frac{1}{2}\int_{0}^{\beta }d\tau \mathbf{b}_{i_{0}}^{*
}(\tau )\left( \partial _{\tau }\boldsymbol{\sigma} _{3}-(\mu _{b}-\epsilon
_{i_{0}}^{b})\mathbf{1}\right) \mathbf{b}_{i_{0}}(\tau )+\frac{1}{2} \int_{0}^{\beta
}d\tau \int_{0}^{\beta }d\tau ^{\prime }\mathbf{b}_{i_{0}}^{* }(\tau )%
\mathbf{\Delta }_{i_{0}}^{b}(\tau -\tau ^{\prime })\mathbf{b}_{i_{0}}(\tau
^{\prime })  \nonumber \\
&&+\frac{U_{b}}{2}\int_{0}^{\beta }n_{i_{0}}^{b}(\tau )(n_{i_{0}}^{b}(\tau
)-1)+\int_{0}^{\beta }d\tau \sum_{j\neq i_{0}}\kappa_{i_{0}j}\mathbf{b}%
_{i_{0}}^{* }(\tau )\mathbf{\Phi }_{j}(\tau )
\label{b1}
\end{eqnarray}%
is the action for bosons coupled to a reservoir of normal and
condensed lattice bosons \cite{byczuk08},
\begin{eqnarray}
S_{i_{0}}^{\mathrm{f}}=\int_{0}^{\beta }d\tau f_{i_{0}}^{\ast }(\tau
)\left(
\partial _{\tau }-\mu _{f}+\epsilon _{i_{0}}^{f}\right) f_{i_{0}}(\tau
)+\int_{0}^{\beta }d\tau \int_{0}^{\beta }d\tau ^{\prime
}f_{i_{0}}^{\ast }(\tau )\Delta _{i_{0}}^{f}(\tau -\tau ^{\prime
})f_{i_{0}}(\tau ^{\prime }) \label{b2}
\end{eqnarray}
is the action for spinless fermions coupled to a reservoir of normal
lattice bosons and
\begin{eqnarray}
 S_{i_{0}}^{\mathrm{bf}}=U_{bf}\int_{0}^{\beta
}d\tau n_{i_{0}}^{b}(\tau )n_{i_{0}}^{f}(\tau )
\end{eqnarray}
is the action describing the coupling between bosons and fermions at site $%
i_{0}$. There is no coupling of fermions to condensed bosons since this
coupling is described by disconnected correlation functions. The symbol $%
\mathbf{1}$ denotes a $2\times 2$ unit matrix, and $\boldsymbol{\sigma}_{3}$ is the Pauli matrices with $\pm 1$  on the diagonal. The
factor $1/2$ is added because in the Nambu representation the number
of degrees of freedom is doubled \cite{rickayzen}. We will now
discuss the physical meaning of the quantities ${\bf
\Delta}^b_{i_0}$, $\Delta _{i_{0}}^{f}$ and $\kappa_{i_{0}j}$
appearing in (\ref{b1}) and (\ref{b2}), and how they are determined.

The quantities $\mathbf{%
\Delta }_{i_{0}}^{b}$ and $\Delta _{i_{0}}^{f}$ describe the resonant
broadening of quantum-mechanical states on lattice site $i_{0}$. They may be
interpreted as a hybridization of bosons and fermions, respectively, on site
$i_{0}$ with the surrounding bosonic or fermionic reservoirs and are given by%
\begin{equation}
\mathbf{\Delta }_{i_{0}}^{b}(i\nu _{n})=-\sum_{ij\neq i_{0}}t_{i_{0}i}^{b}{%
\tilde{\mathbf{G}}}_{ij}^{b}(i\nu _{n})t_{ji_{0}}^{b}\label{d1}
\end{equation}%
and
\begin{equation}
\Delta _{i_{0}}^{f}(i\omega _{n})=-\sum_{ij\neq i_{0}}t_{i_{0}i}^{f}{\tilde{G%
}}_{ij}^{f}(i\nu _{n})t_{ji_{0}}^{f}. \label{d2}
\end{equation}%
Here $\nu _{n}=2n\pi /\beta $ and $\omega _{n}=(2n+1)\pi /\beta $
are even (bosons) and odd (fermions) Matsubara frequencies,
respectively. The tilde denotes cavity Green functions which are
determined for a lattice where the site $i_{0}$ is removed.
Nevertheless, using eqs.~(\ref{d1},\ref{d2}) the
hybridization functions can be expressed by local correlation
functions only. Namely, since the self-energies are diagonal in the
lattice indices
and the cavity Green functions are related to the full Green functions by%
\begin{equation}
{\tilde{\mathbf{G}}}_{ij}^{b}(i\nu _{n})=\mathbf{G}_{ij}^{b}(i\nu _{n})-%
\mathbf{G}_{ii_{0}}^{b}(i\nu _{n})\mathbf{G}_{i_{0}i_{0}}^{b}(i\nu _{n})^{-1}%
\mathbf{G}_{i_{0}j}^{b}(i\nu _{n}),
\end{equation}%
and
\begin{equation}
{\tilde{G}}_{ij}^{f}(i\nu _{n})={G}_{ij}^{f}(i\nu _{n})-{G}%
_{ii_{0}}^{f}(i\nu _{n}){G}_{i_{0}i_{0}}^{f}(i\nu _{n})^{-1}{G}%
_{i_{0}j}^{f}(i\nu _{n}),
\end{equation}%
one can derive \cite{tran07} the corresponding local Dyson equations and express the hybridization functions
by the self-energies and the local Green functions as
\begin{equation}
\mathbf{G}_{i_{0}}^{b}(i\nu _{n})^{-1}+\mathbf{\Sigma }_{i_{0}}^{b}(i\nu
_{n})=i\nu _{n}\boldsymbol{\sigma}_{3}+(\mu _{b}-\epsilon _{i_{0}}^{b})\mathbf{1%
}-\mathbf{\Delta }_{i_{0}}^{b}(i\nu _{n}),
\end{equation}%
and
\begin{equation}
G_{i_{0}}^{f}(i\omega _{n})^{-1}+\Sigma _{i_{0}}^{f}(i\omega _{n})=i\omega
_{n}+\mu _{f}-\epsilon _{i_{0}}^{f}-\Delta _{i_{0}}^{f}(i\omega _{n}).
\end{equation}%
To close the set of the DMFT equations we employ the Dyson equation for the
lattice Green functions, where the exact self-energies are replaced by the
local ones, i.e.,
\begin{equation}
\mathbf{G}_{ij}^{b}(i\nu _{n})=\left[ (i\nu _{n}\boldsymbol{\sigma} _{3}+\mu _{b}%
\mathbf{1}-\mathbf{\Sigma }_{i}(i\nu _{n}))\delta _{ij}-t_{ij}^{b}\mathbf{1}%
\right] ^{-1},  \label{dyson_b}
\end{equation}%
and
\begin{equation}
G_{ij}^{f}(i\omega _{n})=\left[ (i\omega _{n}+\mu _{f}-\Sigma _{i}(i\omega
_{n}))\delta _{ij}-t_{ij}^{f}\right] ^{-1}.  \label{dyson_f}
\end{equation}%
The self-consistent solution is reached when the local Green functions (\ref%
{green_b_loc}) and (\ref{green_f_loc}) are identical to the corresponding
diagonal elements of the Green functions obtained from the Dyson eqs.~(\ref%
{dyson_b}) and (\ref{dyson_f}).

The scaled hopping amplitudes $\kappa_{i_0j}$ appearing in eq. (\ref{b1}) need to be determined such that in the non-interacting ($U_b=0$) case the linear Gross-Pitaevski equation
\begin{equation}
\frac{1}{2}\left( \partial _{\tau }\pmb{\sigma} _{3}-(\mu _{b}-\epsilon
_{i_{0}}^{b})\mathbf{1}\right) \mathbf{\Phi}_{i_{0}}(\tau )+\frac{1}{2} \int_{0}^{\beta }d\tau ^{\prime }
\mathbf{\Delta }_{i_{0}}^{b}(\tau -\tau ^{\prime })\mathbf{\Phi}_{i_{0}}(\tau
^{\prime }) + \sum_{j\neq i_{0}}\kappa_{i_{0}j}\mathbf{\Phi }_{j}(\tau )=0
\end{equation}
is satisfied for a given lattice \cite{werner09}.

The real-space formulation presented here can be employed to study
mixtures of bosons and fermions on a lattice in the presence of
external potentials. It allows for the appearance of phases with
diagonal long-range order, phase separation, and BEC. We note that
when the external potentials are random, one also has to specify the
average over the disorder \cite{aguiar09,byczuk05,byczuk09}. In the absence
of external potentials ($\epsilon _{i}^{b,f}=0$)  there is usually no
need for a formulation in real-space since the lattice Dyson
equations can be written as a Hilbert transform over the momenta.
However, even for $\epsilon _{i}^{b,f}= 0$ a real-space formulation is more suitable when one wants to describe    phase separation or incommensurate long-range order, e.g., stripe formation.

\subsection{Effective attraction between bosons mediated by their interaction
with fermions}

Due to the absence of local fermion--fermion interactions the DMFT action
for model I is bilinear in Grassmann variables. Therefore they can be
integrated out yielding an effective local action for the interacting
bosons. The local partition function reads%
\[
Z_{i_{0}}=\int D[b]e^{-S_{i_{0}}^{\mathrm{b}}[b]+\ln \mathrm{Det}\left[
M_{i_{0}}^{b}\right] },
\]%
where the matrix elements of the operator $M_{i_{0}}^{b}$ in Matsubara
frequency space have the form
\begin{eqnarray}
[M_{i_0}^b]_{nm} &\equiv &\left[ (\partial _{\tau }-\mu _{f}+\epsilon
_{i_{0}}^{f}+U_{bf}n_{i_{0}}^{b}(\tau ))\delta _{\tau ,\tau ^{\prime
}}+\Delta _{i_{0}}^{f}(\tau -\tau ^{\prime })\right] _{nm}  \nonumber \\
&=&\left[ -i\omega _{n}-\mu _{f}+\epsilon _{i_{0}}^{f}+\Delta
_{i_{0}}^{f}(\omega _{n})\right] \delta _{nm}+\frac{U_{bf}}{\sqrt{\beta }}%
n_{i_{0}}^{b}(\omega _{n}-\omega _{m}).
\end{eqnarray}%
Apparently the local interaction between bosons and fermions
$U_{bf}$ gives rise to a complicated effective dynamics of the
bosons. Indeed, even if the interaction between the bosons is
neglected ($U_{b}=0$) the effective \textit{bosonic action} is not
bilinear. We will now show that this corresponds to a nontrivial
effective interaction between the bosons. To this end we expand the
logarithm of the determinant as
\begin{eqnarray}
\ln \mathrm{Det}[M^{b}] &=&\mathrm{Tr}\ln [M^{b}]=\mathrm{Tr}\ln [-(\mathcal{%
G}^{f})^{-1}+M_{1}^{b}]  \nonumber \\
&=&\mathrm{Tr}\ln [-(\mathcal{G}^{f})^{-1}]-\sum_{m=1}^{\infty }\frac{1}{m}%
\mathrm{Tr}[\mathcal{G}^{f}M_{1}^{b}]^{m},
\end{eqnarray}%
where we make use of the Weiss Green function
\begin{equation}
\mathcal{G}_{i_{0}}^{f}(\omega _{n})=\frac{1}{i\omega _{n}+\mu _{f}-\epsilon
_{i_{0}}^{f}-\Delta _{i_{0}}^{f}(\omega _{n})}.
\end{equation}%
By keeping only the first two terms one obtains the Random Phase
Approximation (RPA), which takes into account bubble diagrams for fermions
plus fermionic Hartree contributions. The effective bosonic action is then
found as%
\begin{eqnarray}
\tilde{S}_{i_{0}}^{b}\approx S_{i_{0}}^{b}+\frac{U_{bf}}{\sqrt{\beta }}%
\sum_{n}\mathcal{G}_{i_{0}}^{f}(\omega _{n})n_{i_{0}}^{b}(\nu _{m}=0)-\frac{%
U_{bf}^{2}}{2}\sum_{n}n_{i_{0}}^{b}(\nu _{n})\Pi _{i_{0}}^{f}(\nu
_{n})n_{i_{0}}^{b}(-\nu _{n}),
\end{eqnarray}
where we introduced the fermionic polarization (bubble) function
\begin{equation}
\Pi _{i_{0}}^{f}(\nu _{n})\equiv -\frac{1}{\beta }\sum_{m}\mathcal{G}%
_{i_{0}}^{f}(\omega _{m})\mathcal{G}_{i_{0}}^{f}(\omega _{m}+\nu _{n}).
\end{equation}%
The first-order term in $U_{bf}$ is the Hartree contribution and leads to a
renormalization of the bosonic chemical potential. The second-order term
describes a frequency dependent (retarded) local interaction between bosons
due to presence of the fermions. The static component of the polarization
function is equal to the fermionic local density of states $N_{i_{0}}^{f}$
at the Fermi level, i.e., $\Pi _{i_{0}}^{f}(\nu _{n}=0)=N_{i_{0}}^{f}(\mu )$%
. Therefore the effective static interaction between the bosons is given by $%
U_{b}^{\mathrm{eff}}=U_{b}-U_{bf}^{2}N_{i_{0}}^{f}(\mu )$, which
becomes attractive for $U_{b}<U_{bf}^{2}N_{i_{0}}^{f}(\mu )$. The
bosonic subsystem then becomes unstable against phase separation or
the formation of bosonic molecules. It is interesting to note that
this effective attraction between the bosons originates from their
interaction with the fermions.
This is complementary to the well-known mediation of an effective attraction
between fermions through the exchange of virtual bosons, e.g., phonons. This corroborates earlier results on the effective interaction between bosons  \cite{viverit00,bijlsma00,albus02,buchler03,buchler04,chui104,wang05,rothel07,sengupta07,enss08,lutchyn08,refael08,pollet08,titvinidze08,lutchyn09}.

\subsection{Immobile fermions: Falicov-Kimball limit of model I}

In the case of immobile fermions, i.e., $t_{ij}^{f}=0$ (atomic limit), model
I reduces to a Falicov-Kimball like model \cite{byczuk08,gruber96} for Bose-Fermi
mixtures. The number of fermions on each site is then conserved and can be
used to characterize many-body states. The local partition function takes
the form
\begin{eqnarray}
Z_{i_{0}}=\sum_{n_{i_{0}}=0,1}e^{\beta (\mu _{f}-\epsilon
_{i_{0}}^{f})n_{i_{0}}^{f}}\int
D[b]e^{-S_{i_{0}}^{b}-U_{bf}n_{i_{0}}^{f}\int_{0}^{\beta }d\tau
n_{i_{0}}^{b}(\tau )}.
\end{eqnarray}
Similar expressions can be derived for bosonic correlation
functions. Although the fermions are immobile they are still
thermodynamically coupled to the bosons. This is expressed by the
annealed (rather than quenched) average over different possible
configurations of fermions on the lattice. In other words, due to
the fluctuations around the microscopic equilibrium configurations
the systems probes different configurations\textbf{\ }of fermions
and, for a given temperature and other external parameters, realizes
the optimal configuration. In particular, on a bipartite lattice and
at low enough temperatures the fermions can be long-range ordered
with a checker board structure. Therefore the $t_{ij}^{f}=0$
(Falicov-Kimball) limit of model I differs from an Anderson disorder
model, where fermions are \textit{fixed} at random lattice sites
which are independent of temperature or other parameters.

For $t_{ij}^{f}=0$ model I can have several different solutions. In
particular, a sufficiently strong repulsion between bosons and fermions, $%
U_{bf}$, will cause a splitting of the bosonic band into two
subbands. In this limit, and for a homogeneous system with boson
density $\bar{n}^{b}$ satisfying $\bar{n}^{b}+\bar{n}^{f}\in
\mathbf{\mathbb{N}}$, where $\bar{n}^{f}$ is the density of
fermions, a change of the interaction $U_{b}$ will lead to a
transition from a superfluid to a Mott insulator. Namely, for large
values of $U_{bf}$ the lattice sites occupied by fermions are
inaccessible for bosons, thereby increasing the effective density of
bosons. This transition is an analog of the metal-insulator
transition for fermions at non-integer densities in the presence of
binary-alloy disorder \cite{byczuk04}. Away from these special
densities the
ground state will be a superfluid. Since the condensation temperature $%
T_{BEC}$ increases with the density of bosons we also expect $T_{BEC}$ to
increase with increasing $U_{bf}$. Indeed, for $U_{b}=0$ the model is
equivalent to the bosonic Falicov-Kimball model with immobile, hard-core
bosons discussed in Ref.\cite{byczuk08} where it was shown that the
transition temperature $T_{BEC}$ increases with increasing $U_{bf}$.

\section{Solution of model II within Dynamical Mean-Field Theory}

In the case of Bose-Fermi mixtures with spinful fermions described
by model II, (2), condensation into a state with macroscopic quantum
coherence is not restricted to bosons. Namely, fermions can undergo
\textit{pair condensation} provided $U_{f}<0,$ or if the effective
boson-mediated attraction dominates over the bare repulsion
$U_{f}>0$ (see below). To include the possibility of fermionic
superfluidity (or superconductivity in the case of charged fermions)
within the DMFT we employ the Nambu formalism also for fermions. To
this end we define Nambu spinor fermion operators
\begin{equation}
\mathbf{f}_{i}=\left(
\begin{array}{c}
f_{i\sigma } \\
f_{i\bar{\sigma}}^{\dagger }%
\end{array}%
\right) \;\;\;\mathrm{and}\;\;\;\mathbf{f}_{i}^{\dagger }=(f_{i\sigma
}^{\dagger },f_{i\bar{\sigma}})
\end{equation}%
and the local fermionic one-particle Green function
\begin{equation}
\mathbf{G}_{i}^{f}(\tau )\equiv -\langle T_{\tau }\mathbf{f}_{i}(\tau )%
\mathbf{f}_{i}^{\dagger }(0)\rangle _{S_{i}}\equiv -\left(
\begin{array}{cc}
\langle T_{\tau }f_{i\sigma }(\tau )f_{i\sigma }^{\dagger }(0)\rangle
_{S_{i}} & \langle T_{\tau }f_{i\sigma }(\tau )f_{i\bar{\sigma}}(0)\rangle
_{S_{i}} \\
\langle T_{\tau }f_{i\bar{\sigma}}^{\dagger }(\tau )f_{i\sigma }^{\dagger
}(0)\rangle _{S_{i}} & \langle T_{\tau }f_{i\bar{\sigma}}^{\dagger }(\tau
)f_{i\bar{\sigma}}(0)\rangle _{S_{i}}%
\end{array}%
\right) .  \label{green_f_loc_spin}
\end{equation}%
For bosons all relevant Green functions and the BEC order parameter are the
same as in model I.

To derive the DMFT equations for model II in real space we again
employ the cavity method and proceed as before. As a result we find
the following local DMFT action at site $i_{0}$
\begin{equation}
S_{i_{0}}=S_{i_{0}}^{\mathrm{b}}+S_{i_{0}}^{\mathrm{f}}+S_{i_{0}}^{\mathrm{bf%
}},
\end{equation}%
where
\begin{eqnarray}
S_{i_{0}}^{\mathrm{b}} &=&\frac{1}{2}\int_{0}^{\beta }d\tau \mathbf{b}_{i_{0}}^{*
}(\tau )\left( \partial _{\tau }\pmb{\sigma} _{3}-(\mu _{b}-\epsilon
_{i_{0}}^{b})\mathbf{1}\right) \mathbf{b}_{i_{0}}(\tau )+\frac{1}{2}\int_{0}^{\beta
}d\tau \int_{0}^{\beta }d\tau ^{\prime }\mathbf{b}_{i_{0}}^{* }(\tau )%
\mathbf{\Delta }_{i_{0}}^{b}(\tau -\tau ^{\prime })\mathbf{b}_{i_{0}}(\tau
^{\prime })  \nonumber \\
&&+\frac{U_{b}}{2}\int_{0}^{\beta }n_{i_{0}}^{b}(\tau )(n_{i_{0}}^{b}(\tau
)-1)+\int_{0}^{\beta }d\tau \sum_{j\neq i_{0}}\kappa_{i_{0}j}\mathbf{b}%
_{i_{0}}^{* }(\tau )\mathbf{\Phi }_{j}(\tau )
\end{eqnarray}%
is the action for bosons coupled to a reservoir of normal and
condensed lattice bosons\cite{byczuk08},
\begin{eqnarray}
S_{i_{0}}^{\mathrm{f}} &=&\int_{0}^{\beta }d\tau \mathbf{f}_{i_{0}}^{\ast
}(\tau )\left( \partial _{\tau }\mathbf{1}-(\mu _{f}-\epsilon _{i_{0}}^{f})%
\pmb{\sigma} _{3}\right) \mathbf{f}_{i_{0}}(\tau )+\int_{0}^{\beta }d\tau
\int_{0}^{\beta }d\tau ^{\prime }\mathbf{f}_{i_{0}}^{\ast }(\tau )\mathbf{%
\Delta }_{i_{0}}^{f}(\tau -\tau ^{\prime })\mathbf{f}_{i_{0}}(\tau ^{\prime
}),  \nonumber \\
&&+\frac{U_{f}}{2}\int_{0}^{\beta }\sum_{\sigma} n_{i_{0}\sigma}^{f}(\tau
)n_{i_{0}\bar{\sigma}}^{f}(\tau )
\end{eqnarray}%
is the action for spinful fermions coupled to a fermionic reservoir,
and
\begin{eqnarray}
S_{i_{0}}^{\mathrm{bf}}=U_{bf}\int_{0}^{\beta }d\tau \sum_{\sigma
}n_{i_{0}}^{b}(\tau )n_{i_{0}\sigma }^{f}(\tau )
\end{eqnarray}
is the action describing the coupling between bosons and fermions at site $%
i_{0}$. The hybridization function matrices $\mathbf{\Delta }_{i_{0}}^{b}$
and $\mathbf{\Delta }_{i_{0}}^{f}$ are related to the local self-energies
through the local Dyson equations
\begin{equation}
\mathbf{G}_{i_{0}}^{b}(i\nu _{n})^{-1}+\mathbf{\Sigma }_{i_{0}}^{b}(i\nu
_{n})=i\nu _{n}\pmb{\sigma} _{3}+(\mu _{b}-\epsilon _{i_{0}}^{b})\mathbf{1%
}-\mathbf{\Delta }_{i_{0}}^{b}(i\nu _{n}),
\end{equation}%
and
\begin{equation}
\mathbf{G}_{i_{0}}^{f}(i\omega _{n})^{-1}+\mathbf{\Sigma }%
_{i_{0}}^{f}(i\omega _{n})=i\omega _{n}\mathbf{1}+(\mu _{f}-\epsilon
_{i_{0}}^{f})\pmb{\sigma} _{3}-\mathbf{\Delta }_{i_{0}}^{f}(i\omega _{n}).
\end{equation}%
The set of self-consistent DMFT equations for the model II is closed by the
Dyson equations for the bosonic and fermionic lattice Green functions with
local self-energies
\begin{equation}
\mathbf{G}_{ij}^{b}(i\nu _{n})=\left[ (i\nu _{n}\pmb{\sigma} _{3}+\mu _{b}%
\mathbf{1}-\mathbf{\Sigma }_{i}(i\nu _{n}))\delta _{ij}-t_{ij}^{b}\mathbf{1}%
\right] ^{-1},  \label{dyson_b_II}
\end{equation}%
and
\begin{equation}
\mathbf{G}_{ij}^{f}(i\omega _{n})=\left[ (i\omega _{n}\mathbf{1}+\mu _{f}%
\pmb{\sigma} _{3}-\mathbf{\Sigma }_{i}(i\omega _{n}))\delta
_{ij}-t_{ij}^{f}\mathbf{1}\right] ^{-1}.  \label{dyson_f_II}
\end{equation}%
These equations yield the general DMFT solution for Bose-Fermi mixture on a
lattice. They include the possibility for bosonic and fermionic
superfluidity, phase separation, and the appearance of phases with diagonal
long-range order.

\subsection{Effective attraction between fermions mediated by the
interaction with bosons}

In model II the presence of fermions with a local interaction can
lead to similar effects as those discussed in connection with model
I, namely to an effective attraction between bosons, and band
splitting in the Falicov-Kimball (atomic) limit where
$t_{ij}^{f}=0$. In particular, we will now discuss the situation
where bosons influence the effective local interaction between
fermions. For non-interacting bosons, $U_{b}=0$, the local action is
bilinear in the bosonic field which can therefore be integrated out
analytically. We introduce a Bogoliubov shift of the complex
bosonic field to remove the linear term in the bosonic part of the
action as
\begin{equation}
b(\tau )\rightarrow \tilde{b}(\tau )+c(\tau ),\;\;b^{\ast }(\tau
)\rightarrow \tilde{b}^{\ast }(\tau )+c^{\ast }(\tau ).
\end{equation}%
Here $\tilde{b}(\tau )$ and $c(\tau )$ represent normal and condensed
bosons, respectively. The local bosonic action then takes the form
\begin{eqnarray}
S_{i_{0}}^{b} &=&\int_{0}^{\beta }d\tau \tilde{b}_{i_{0}}^{\ast }(\tau
)\left( \partial _{\tau }-\mu _{b}+\epsilon _{i_{0}}^{b}+U_{bf}\bar{n}%
_{i_{0}}^{f}\right) \tilde{b}_{i_{0}}(\tau )+\int_{0}^{\beta }d\tau
\int_{0}^{\beta }d\tau ^{\prime }\tilde{b}_{i_{0}}^{\ast }(\tau )\Delta
_{i_{0}}^{b}(\tau -\tau ^{\prime })\tilde{b}_{i_{0}}(\tau )  \nonumber \\
&&+\frac{\sum_{j\neq i_{0}}\kappa_{i_{0}j}|\phi _{j}|}{\mu _{b}-\epsilon
_{i_{0}}^{b}-U_{bf}\bar{n}_{i_{0}}^{f}-\Delta _{i_{0}}^{b}(\nu _{n}=0)},
\end{eqnarray}%
where the last term describes the Bose-Einstein condensate, and the chemical
potential for bosons is renormalized due to the interaction with the
fermions by a Hartree-like term. The contribution describing the interaction
between bosons and fermions is now expressed in terms of normal bosons $%
\tilde{b}$, i.e.,
\begin{equation}
S_{i_{0}}^{bf}=U_{bf}\int_{0}^{\beta }d\tau \tilde{b}_{i_{0}}^{\ast }(\tau )%
\tilde{b}_{i_{0}}(\tau )n_{i_{0}}^{f}(\tau ).
\end{equation}%
In the next step we integrate out the normal bosons and thereby obtain the
following partition function for fermions:
\begin{eqnarray}
Z_{i_{0}}^{f}=\int D[f]e^{-S_{i_{0}}^{f}[f]-\ln \mathrm{Det}\left[
M_{i_{0}}^{f}\right] }.
\end{eqnarray}
Here the matrix elements of the operator $M_{i_{0}}^{f}$ in
Matsubara frequency space have the form
\begin{eqnarray}
\left[ M_{i_{0}}^{f}\right] _{nm} &=&\left[ (\partial _{\tau }-\mu
_{b}+U_{bf}\bar{n}_{i_{0}}^{f}+\epsilon _{i_{0}}^{b}+U_{bf}\sum_{\sigma
}n_{i_{0}\sigma }^{f})\delta _{\tau \tau ^{\prime }}+\Delta
_{i_{0}}^{b}(\tau -\tau ^{\prime })\right] _{nm}  \nonumber \\
&=&\left[ -i\nu _{n}-\mu _{b}+U_{bf}\bar{n}_{i_{0}}^{f}-\epsilon
_{i_{0}}^{b}+\Delta _{i_{0}}^{b}(\nu _{n})\right] \delta _{nm}+\frac{U_{bf}}{%
\sqrt{\beta }}\sum_{\sigma }n_{i_{0}\sigma }^{f}(\nu _{n}-\nu _{m}).
\end{eqnarray}%
Finally we introduce the Weiss Green function for local bosons
\begin{equation}
\mathcal{G}_{i_{0}}^{b}(\nu _{n})=\frac{1}{i\nu _{n}+\mu _{b}-U_{bf}\bar{n}%
_{i_{0}}^{f}-\epsilon _{i_{0}}^{b}-\Delta _{i_{0}}^{b}(\nu _{n})},
\end{equation}%
and perform a Taylor expansion to second order in $U_{bf}$. The effective
fermionic action then takes the form
\[
\tilde{S}_{i_{0}}^{f}\approx S_{i_{0}}^{f}-\frac{U_{bf}}{\sqrt{\beta }}%
\sum_{n\sigma }\mathcal{G}_{i_{0}}^{b}(\nu _{n})n_{i_{0}}^{f}(0)+\frac{%
U_{bf}^{2}}{2}\sum_{n\sigma \sigma ^{\prime }}n_{i_{0}\sigma
}^{f}(\nu _{n})\Pi _{i_{0}}^{b}(\nu _{n})n_{i_{0}\sigma ^{\prime
}}^{f}(-\nu _{n}),
\]%
where the bosonic polarization function is given by
\begin{equation}
\Pi _{i_{0}}^{b}(\nu _{n})=\frac{1}{\beta }\sum_{m}\mathcal{G}%
_{i_{0}}^{b}(\nu _{m})\mathcal{G}_{i_{0}}^{b}(\nu _{m}+\nu _{n}).
\end{equation}%
In the static limit this polarization function yields the local
compressibility, $\Pi _{i_{0}}^{b}(\nu _{n}=0)=-\partial \bar{n}%
_{i_{0}}^{b}/\partial \mu _{b}$, which must be positive for the
system to be stable. The effective local interaction between
fermions $U_f^{eff} = U_f - U_{bf}^2 \partial \bar{n}%
_{i_{0}}^{b}/\partial \mu _{b}$ becomes
negative (attractive) for $U_f < U_{bf}^2 \partial \bar{n}%
_{i_{0}}^{b}/\partial \mu _{b}$. Hence a
repulsive interaction between fermions and bosons can lead to an
effective attraction between fermions. The fermionic subsystem then
becomes unstable with respect to a condensation of Cooper pairs as
is well-known from the theory of superconductivity \cite{ashcroft}.

\section{Summary}

We derived a comprehensive, thermodynamically consistent theoretical
framework for the investigation of mixtures of correlated lattice
bosons and fermions --- the BF-DMFT. In analogy to its purely
fermionic and bosonic counterparts the BF-DMFT becomes exact in the
limit of high spatial dimensions $d$ or coordination number $Z$. It
may be employed to calculate the phase diagram and thermodynamics of
mixtures of interacting lattice bosons and fermions in the entire
range of microscopic parameters. As in the B-DMFT the BF-DMFT
employs a different scaling of the hopping amplitude of bosons with
$Z$, depending on whether  the bosons are in the normal or the
condensed phase. Thereby normal and condensed bosons are treated on
equal footing. Namely, in the BF-DMFT the normal bosons retain their
dynamics even in the limit $Z\rightarrow \infty$, such that the
effects of the dynamic coupling between normal bosons and the
condensate are fully included.

Using the self-consistency equations of the BF-DMFT we solved two
different interaction models of correlated bosons and fermions:
model I where all particles are spinless, and model II where
fermions carry spin one-half. In the latter model a local
interaction between fermions is present which leads to dynamical
effects even in the limit of large $Z$. In model I we showed that
the local interaction between bosons and fermions can give rise to
an effective attraction between bosons. The bosonic subsystem is
therefore unstable against phase separation or the formation of
bosonic molecules. In the case of immobile fermions model I reduces
to a Falicov-Kimball model for Bose-Fermi mixtures which can be
solved in closed form. Model II has several different solutions
depending on the interaction parameters and the density of bosons. In
particular, the interaction between bosons and fermions can mediate
an effective attractive local interaction not only between bosons
but also between fermions. The latter leads to an instability with
respect to Cooper pair formation as in superconductivity.

In conclusion, interaction effects in mixtures of bosons and
fermions on a lattice can lead to a multitude of fascinating
correlation phenomena, including the emergence of an effective
attraction between bosons mediated by fermions and vice versa.
Experiments with cold atoms in optical lattices will be able to test
these predictions and lead to deep insights into the overall
physical behavior of many-body systems with particles obeying
different statistics.

\begin{acknowledgement}
We thank A. Kampf, A. Kauch and P. Werner for many useful
discussions. This work was supported in part by the
Sonderforschungsbereich 484 of the Deutsche Forschungsgemeinschaft
(DFG). KB also acknowledges a grant of the Polish Ministry of
Science and Education N202026 32/0705.

\end{acknowledgement}


\begin{thebibliography}{99}
\bibitem{NN_interactions} There is, of course, an interaction between fermions
on nearest-neighbor sites. Since the van-der-Waals interaction
between free atoms is short range and the lattice Wannier wave
functions are localized this interaction is neglected with respect
to all local interactions. In addition,  in the DMFT approach
employed here such an interation reduces to static, Hartree-type
potentials which merely renormalize the chemical potential
\cite{mh89}.
\bibitem{Bosonic_scaling} As discussed in Ref.\cite{byczuk08} a
comprehensive bosonic dynamical mean-field theory (B-DMFT), which
treats normal and condensed bosons on equal footing and becomes
exact in the limit of infinite dimensions $d$\ or coordination
numbers $Z$, is obtained by
scaling the hopping amplitudes as $t_{ij}^{b}=\tilde{t}%
_{ij}^{b}/(Z^{R_{ij}})^{s}$, where $R_{ij}$ is the distance between
sites $i$ and $j$, with $s=1$ if the bosons are quantum condensed
and $s=1/2$ if they are in the normal state, respectively.
\bibitem{comment_a} The notation
$\langle T_{\tau} b(\tau)b^{\dagger}(\tau ')\rangle_{S} $, where
$b(\tau)$, $b^{\dagger}(\tau)$ are bosonic operators in the
Heisenberg representation, is meant to imply that the quantum
mechanical and statistical averages are calculated within a path
integral formalism with the action $S$. Namely, $\langle T_{\tau}
b(\tau)b^{\dagger}(\tau ')\rangle_{S} =  \int D[b]b(\tau) b^*(\tau
') \exp(-S[b])/Z$, where the functional integral is performed in the
space of complex  functions $b(\tau)$ which obey periodic boundary
conditions $b(\tau+\beta)=b(\tau)$, and $Z=\int D[b] \exp(-S[b])$
\cite{negele}. In the case of fermions the corresponding functional
integrals are performed with Grassmann (anticommuting) variables
satisfying anti-periodic boundary conditions.
\end{thebibliography}
\end{document}